# Promoting target models by potential measures

Dr. Jörg Dubiel, Scientific-Marketing e.V.

**Abstract**: Direct marketers use target models in order to minimize the spreading loss of sales efforts. The application of target models has become more widespread with the increasing range of sales efforts. Target models are relevant for offline marketers sending printed mails as well as for online marketers who have to avoid intensity. However business has retained its evaluation since the late 1960s. Marketing decision-makers still prefer managerial performance measures of the economic benefit of a target model. Such benefit measures have merits but are unfavorable in other respects: They constrain leadership by stretched targets since they do not tell us how good a model could be. And they require a predisposed decision regarding cut-offs. Since this is based on earlier optimizations it virtually means sticking to traditions. Hence it is recommended also to use cut-off invariant and potential oriented performance measures for the model evaluation. This has three advantages: sustaining stretched targets, identifying improvement potential and supporting an automated evaluation of many different models. This article introduces a concrete potential measure and shows how to calculate it. It is especially recommended for direct marketing businesses churning out many specific target models at short intervals.

## 1  Introduction of terms: score models for minimizing spreading loss in sales promotion

We explain the core terms of the article using the example of a Direct Mail company. Direct Mail companies use expensive promotion material in order to generate revenues or in more abstract terms, to stimulate a desired behavior of the recipient. Promotion material which generates no revenues or does not stimulate the behavior of preference is considered a spreading loss of sales promotion. Even in the era of online business spreading loss is a relevant criterion. Presumably, direct marketers who address premium customer target groups will increasingly (Moestel, 2010) use printed mails. The reason for this is to create some distinction from online business by expressing an implicit value, as the sales promotion is a haptic and therefore sensually perceivable experience. The spreading loss is calculated as (costs per direct mail action) minus (costs per responder). The costs of a direct mail action i.e. a mailing can be calculated according to Shepard (Shepard, 1999, p. 269) as: costs per thousand: total costs of direct marketing action divided by the number of addresses multiplied by 1000. The costs per responder i.e. the orderers can be calculated according to Brann (Brann, 1984, p. 19) as costs per thousand: total costs of direct marketing action divided by the number of responders per thousand. In order to minimize this spreading loss, advanced direct marketing companies use mathematical models, so-called score-models or "targeting models" (Nash, 1993, p. 133). The first of these models  - based on discriminant analysis and regression analysis – were successfully tested and used by pioneering companies like Readers Digest in 1969 (Nash, 1993, p. 140).

For each customer, these models calculate a value. The "score value" often is a linear combination of customers' data, still often in form of a regression technique (Nash, 1993, p. 133). Usually score models are developed on sufficient tests that include real life data from a field test. The sufficient level of significance for direct-marketing tests amounts to a traditional 5% (Stone, 1989, p. 478) based on a sufficient sample-size (Knauff, 1991, p. 583). Tests are carried out in accordance with the steps of effective marketing research (Huldi, 1992, p. 193; Kotler, 1989, pp. 640ff.; Roberts & Berger, 1989, p. 189). Tests are the "supreme imperative of a direct marketer" (Bird, 1990, p. 367) because it is „regardless of all expert –knowledge virtually impossible, to forecast the results of a direct marketing sales campaign – no matter which type" (Bird, 1990, p. 367). Beside that -not just for analytical reasons but also w.r.t quality management - tests are important for checking the quality of outsourced services (Brown & Buskirk, 1989, p. 129).

The resultant score model calculates a score value for each single entity in the sample (for instance a customer). This score value indicates how likely that entity is to show the designated behavior upon receiving the sales promotion. The higher the score values, the higher the likelihood of showing the desired behavior. Once the score model has been developed it can be used to calculate a score value for each entity within the rest of the notionally promotable potential. This is the closed loop concept of scoring, since "testing is not an isolated marketing research activity and should not be treated as such"(Stone, 1989, pp. 310ff.)

Consequently one can discriminate those customers who are worth being promoted, since they are likely to show the desired behavior (pass names) from those who are not (fail names), and thus save promotion money and minimize potential irritation (a negative attitude towards addressed direct mail is approved by a study of PTT (PTT, SVD: Schweizerischer Verband für Direktmarketing, Generaldirektion, 1989, pp. Schweizerischer Verband für Direktmarketing, Generaldirektion 1989 #10: 9). The ratio of pass names to fail names is often called "cut-off". By implication, 10 "pass-names" and 90 "fail names" from a potential of 100 names equals a cut-off of 10%.

## 2   Benefit- and potential-oriented performance measures of score models

An efficient direct marketing business can save a lot of money and thus generate a lot more profit by minimizing spreading loss of sales promotion: "It is an industry axiom that a poor mailing to a good list can be profitable, but that no mailing - no matter how well conceived – will work, if sent to the wrong list" (Nash, 1986, p. 97). This is even true if a sales promotion does not cost much. That is typically the case for electronic direct mails sent via internet. If an electronic direct mail is transmitted to the wrong recipients it is obvious that not a lot of promotion money has been involved. However it will be costly for the sender's reputation and lower the acceptance of subsequent electronic direct mails from the same company. Therefore even such an apparently inexpensive spreading loss costs money. The more an electronic direct mail is customized the more a dis-aimed one will produce spreading-loss.

The principal interest therefore will be to attain and maintain a high standard of the score model's performance. The question however is: How can the quality of a score model be measured? The developers of score models as skilled statistical analysts like using "technical" test sta-tistics (Nash, 1993, p. 155). Marketing managers responsible for P&L decisions prefer using managerial performance measures in order to quantify an easy-to-understand and business-relevant economic contribution of score models (Nash, 1993, pp. 156f.). The term measure will henceforth be used from that managerial perspective.

There are many different approaches to measuring the quality of a score model. In our case, two main groups of measures can be distinguished. This implies "benefit-oriented" and "potential-oriented" performance measures.

Typically, most measures are benefit-oriented. For instance Wilde (Wilde & Hicketier E, 1997, pp. 478ff.) describes different classes of Performance Measures. All of them are benefit oriented in the sense of measuring an absolute benefit of the score model. Although Wilde uses the term "Potential for Success Approaches" (Wilde & Hicketier E, 1997, p. 483), it refers to the potential of a customer or market, not to the potential of a score model.

Typically most measures represent the absolute benefit of that model, by response rate (Nash, 1993, pp. 156f.; Holland, 2009), ROI (Huldi, 1992, p. 234; Nash, 1986, pp. 395f.), marginal income (Ruhland, 1994), customer lifetime value (Keane & Wang, 1995, pp. 59ff.; Shaw & Stone, 1988, p. 136; Roberts & Berger, 1989, p. 409) etc. All these measures reflect what a score model contributes to the chosen criteria of benefit given a certain cut-off. It does not tell us, however, how good it could be independent of a certain cut-off. Benefit-oriented performance measures do not show a relative but an absolute benefit contribution which dependents on a certain cut-off. This has two major disadvantages:

2.1    Disadvantages of benefit-oriented performance measures

Firstly: The quality management of benefit-oriented performance measures therefore typically revolves around two aspects: A, the expected appreciation of the score model's added value, the score gain: is it sufficient to satisfy internal prospects? B, the model's reliability: does it suffice to accomplish the test-based predicted score gain in a practical application?

Secondly. The cut-off dependency usually leads to a suboptimality. Nash describes that effect as follows: "It is the names that respond at a rate near break-even where increased sensitivity in the regression model is most beneficial since the names that respond at a rate significantly above break-even are going to be mailed whether they score a little higher or lower, and conversely, the names that respond at a rate significantly below break-even will not be selected for promotions even if their score changes a little" (Nash, 1993, p. 156). The point of break-even tells us how many orders are needed to make some profit by the sales campaign in question Schaller 1988 #13: 139} meaning a certain-cut off. An example of how to calculate the break-even cut-off can be found in (Huldi, 1992, pp. 230ff.) and (Schaller, 1991, pp. 593ff.). Nash (Nash, 1993, p. 156) solves that issue with a Break-Even Regression, which is a two state regression-methodology. The second regression requires a-priori information about a cut-off. Hence it is still a cut-off dependent method: "This results in a regression model that is most accurate in predicting the customers response rate near break-even level and therefore selects the group of customers with the greatest overall response rate." (Nash, 1993, p. 156) Hence: different cut-offs will generate vari-ous "most accurate" models. But if the final cut-off is a priori not known (i.e. due to unknown break-even criteria) a Break-Even regression will not suffice.

2.2   Advantages of potential-oriented measures

In contrast to benefit-oriented performance measures, a potential-oriented one gives an idea of what a score model *could* achieve. It does not identify an absolute benefit, but delivers a fraction of the potential benefit the model can achieve. It could be interpreted as the prognostic-efficiency of a score model. Having a high standard of prognostic-efficiency also implies a high level of benefit contribution, because a maximum of prognostic-efficiency always means a maximum of benefit. This directs the analyst's – or their manager's - attention to the question: how far are we away from the best model and how could we become better? This leads to the psychological aspects of goal-setting. A potential oriented measure implies a latent lack of performance by taking the ideal solution as the benchmark. This stretching of targets leads to better employee performance (Sherman, 1995), (Tully S., 1994), (Cummings & Worley, 2005, p. 396). Furthermore a potential-oriented performance measure does not depend on a certain cut-off (see the following section 3). Therefore it does not suffer from a non-optimized a priori cut-off decision.

## 3   Transforming to a potential-oriented measure of performance

The following subsection presents a practical example of the migration from a widely used benefit-oriented performance measure to a potential-oriented case. Let us first consider the following benefit-oriented index herein referred to as "BenI".

3.1   The benefit-oriented score measure BenI

To illustrate the calculation of the score measure BenI, we can assume the following table of values:

| Table 1: Exemplary Values | | |
|---|---|---|
| Abbr. | Denotation | Value |
| X | is the size of a test sample (a cross-section) containing data from a field test i.e. response data of a specific sales promotion | 100 |
| R | is the raw response rate generated by the test sample X above | 8% |
| T | is the size of a "total potential", meaning the available number of persons we could promote with the specific sales promotion | 100.000 |
| P | is the number of "pass names", those names which met the threshold of a score model | 4.000 |
| R (P) | is the optimized response rate we generate with pass names | 15 % |
| Cut-Off | a threshold value that determines ratio of admissible names for promotion | 40 % |

Based on these values, we can calculate BenI as a measure of the performance of a score model as follows:

(1) Eq. 1: $\text{BenI} = \dfrac{R(P)}{R(X)} \cdot 100 = \dfrac{15}{8} \cdot 100 = 187$

In business terminology users say "the score model has an index of 187". This means the 40% best "scored" pass names have a 1.87 times better response rate than randomly selected "un-scored" names from the test sample. Therefore BenI is a benefit index that quantifies the benefit contribution. It is not a potential index as it does not tell us how good the model is w.r.t an optimal model (even setting the benefit index BenI in relation to a maximum benefit index is likewise not sufficient to serve as potential index as it is shown in the explanation below for column 6).

For example: the maximum BenI-value based on a given cut-off, can be calculated as

(2) Eq. 2: $B_{max} = \dfrac{100}{cut-off}$ ; (whereby R(X) < cut-off)

Inserting the values from the table above follows

(3) Eq. 3: $B_{max} = \dfrac{100}{0,4} = 250$

The table 2 below shows the marginal values of BenI per bucket in col. 3 and cumulated in col. 4.

Col. 5 calculates the theoretic maximum value of BenI per bucket according to the above equation 2.

Col. 6 yields the cumulative ratio of (BenI max/BenI) per bucket which sums up to 100% and is therefore cut-off dependent. For a 40% cut-off as an example, the ratio tells us that the specific score model attains 75% of the theoretic maximum in that bucket. It does not tell us from an overall and therefore cut-off independent perspective, how far it is from an expected maximum, because it always adds up to 100%.

Table 2:   Distribution of Benefit-Index BenI in a gains-chart

Parameters
No. of Buckets          #B          10
Size of sample X        #X          100
Responsrate in sample X R(X)        8%

| Col. 1 | 2 | 3 | 4 | 5 | 6 |
|---|---|---|---|---|---|
| Cut-Off | ⊕ # Resp. | BenI, marg. | BenI, cum. | BenI, max, cum. | BenI max / BenI |
| 10% | 3 | 375 | 375 | 1000 | 38% |
| 20% | 1 | 125 | 250 | 500 | 50% |
| 30% | 1 | 125 | 208 | 333 | 63% |
| 40% | 1 | 125 | 188 | 250 | 75% |
| 50% | 0 | 0 | 150 | 200 | 75% |
| 60% | 0 | 0 | 125 | 167 | 75% |
| 70% | 1 | 125 | 125 | 143 | 88% |
| 80% | 1 | 125 | 125 | 125 | 100% |
| 90% | 0 | 0 | 111 | 111 | 100% |
| 100% | 0 | 0 | 100 | 100 | 100% |

3.2   The cut-off independent and potential-oriented performance measure Score Potential

As an optimal cut-off (%) definitely varies with different suboptimal score models (the best cut-off of an optimal model equals the raw response rate R), it is advantageous to have an "all-purpose" measure which is %-independent. With such a %-independent measure one can choose the best % *after* optimization of the score model. Such a cut-off invariant performance measure is referred to as Score Potential.

Digression: If an optimal cut-off is exclusively considered, one could measure the model's efficiency just at the point of an optimal %=R. The ratio (BenI/BenI max) could then serve as potential-oriented performance measure The best model is that with a 100% in col. 6 on a % = R.

A definite calculation of Score Potential requires a ranking of all names in the test sample by score values. For each name we assign a rank number x whereby "1" represents the name with the lowest score value and "#X" the name with the highest score value (#X is the size of test sample X). Furthermore, each name $N_x$ from sample X is assigned a response value $V_x$, which is either "1" if $N_x$ is a responder or "0" in case $N_x$ is a non-responder. See table 3 below for an example.

Based on the values in table 3, Score Potential (in the following also abbreviated wit PoP: pro-

Table 3: Exemplary values to calculate the benefit measure PoP

| Pos No. x | Score value | Responder | V | x * V |
|---|---|---|---|---|
| 10 | 4000 | no | 0 | 0 |
| 9 | 3999 | yes | 1 | 9 |
| 8 | 3031 | no | 0 | 0 |
| 7 | 2900 | no | 0 | 0 |
| 6 | 2500 | yes | 1 | 6 |
| 5 | 2455 | yes | 1 | 5 |
| 4 | 2100 | no | 0 | 0 |
| 3 | 1900 | no | 0 | 0 |
| 2 | 1600 | no | 0 | 0 |
| 1 | 500 | no | 0 | 0 |
| | | | | 20 |

portion of potential) can be calculated as

(4) Eq. 4: $$\text{PoP} = \frac{P\uparrow}{P\downarrow} = \frac{\sum_{x=1}^{\#X}(x \cdot V_x)}{-\left(\frac{X^2 R^2}{2} - \left(X^2 + \frac{X}{2}\right) \cdot R\right)} \cdot 100$$

Putting in the values of table 3, Score Potential amounts to

(5) Eq. 5: $$\text{PoP} = \frac{20}{10+9+8} \cdot 100 = 74\% \ .$$

## 3.3 Combined report of BenI and Score Potential

The calculation of Score Potential for an automatic model evaluation can easily be conducted by means of the procedure described above. Besides Score Potential's usage for a programmed model evaluation, it may prove advantageous to calculate Score Potential also in case of manual review of few score models. For example just to determine how far the analyst is away from the perfect solution with one comparable figure. In case of manual model review an approximate Score Potential-value can be calculated within a traditional gains chart (Nash, 1993, pp. 156f.) which spreads the test sample into different buckets e.g. deciles. Now we can first assign a specific value of BenI to each bucket using the above mentioned equation 1 with the bucket-specific values. A basic example is in table 2 above. Score Potential can now be approximately calculated as shown in table 4 below (consider that some parameters and distributions in table 4 have changed as opposed to table 2).

**Table 4: Combined presentation & calculation of BenI and PoP in a gains-chart**

Parameters
No. of Buckets           #B      10
Size of sample X         #X      100
Responsrate in sample X  R(X)    4%

| Col 1 | 2 | 3 | 4 | 5 | 6 | 7 | 8 | 9 | 10 | 11 |
|---|---|---|---|---|---|---|---|---|---|---|
| Bucket No | # Resp. ⊕ | P'↑, max | P'↑, min | P'↑ avg | PoP', marg | PoP', cum. | BenI', marg | BenI', cum | BenI', cum, max | Col 9 / Col 10 |
| 10 | 0 | 0 | 0,0 | 0 | 0% | 0% | 0 | 0 | 1000 | 0% |
| 9 | 0 | 0 | 0,0 | 0 | 0% | 0% | 0 | 0 | 500 | 0% |
| 8 | 3 | 23,7 | 21,6 | 22,65 | 57% | 57% | 750 | 250 | 333 | 75% |
| 7 | 1 | 7 | 6,1 | 6,55 | 17% | 74% | 250 | 250 | 250 | 100% |
| 6 | 0 | 0 | 0,0 | 0 | 0% | 74% | 0 | 200 | 200 | 100% |
| 5 | 0 | 0 | 0,0 | 0 | 0% | 74% | 0 | 167 | 167 | 100% |
| 4 | 0 | 0 | 0,0 | 0 | 0% | 74% | 0 | 143 | 143 | 100% |
| 3 | 0 | 0 | 0,0 | 0 | 0% | 74% | 0 | 125 | 125 | 100% |
| 2 | 0 | 0 | 0,0 | 0 | 0% | 74% | 0 | 111 | 111 | 100% |
| 1 | 0 | 0 | 0,0 | 0 | 0% | 74% | 0 | 100 | 100 | 100% |

P↑           =            29,2
P↓= 10+9,9+9,8  =         39,4
PoP=P↑/P↓    =            74%

An explanation to table 4:

Column 1 shows the position number of the score-ranked buckets (reminder: the bucket no. 1 labels the bucket with the lowest scores. The highest bucket no. labels the bucket with the highest scores).

Column 2 represents the number of responders (⊕) within each bucket.

Column 3 calculates the maximum value of P↑ per bucket (P↑') assuming that all responders are shifted to the top of the bucket. For example the maximum value of the 8$^{th}$ bucket equals 8+7.9+7.8=23.7. That value can be calculated using an arithmetic row with the maximum initial value and descending order as

(6) Eq. 6: $\quad P'\uparrow \max = BNo * (\#\oplus) + \left(-\dfrac{\#B}{\#X}\right) * \dfrac{((\oplus - 1) * \oplus)}{2}$

For instance, considering the row of the 8$^{th}$ bucket in table 4 above we have the following values:

$BNo$ = 8      (see column 1, Bucket No)
$\#\oplus$ = 3      (see column 2, # Resp. or ⊕)
$\#B$ = 10     (see parameters: No. of buckets)
$\#X$ = 100    (see parameters: Size of sample)

Which yields P'↑ in the 8$^{th}$ bucket as: $P'\uparrow \max = 8*3 + \left(\dfrac{-10}{100}\right) * \dfrac{((3-1)*3)}{2} = 23.7$

Column 4 calculates the minimum value of P2↑ per bucket (for instance 8$^{th}$ bucket: 7.1+7.2+7.3=21.6) likewise as an arithmetic row, but now with the minimum initial value and an ascending row as

(7) Eq. 7: $P'\uparrow \min = \left(BNo - 1 + \dfrac{\#B}{\#X}\right) * (\#\oplus) + \left(\dfrac{\#B}{\#X}\right) * \dfrac{((\oplus - 1) * \oplus)}{2}$

Column 5 calculated the average value of P2↑, approx. as

(8) Eq. 8: $P'\uparrow approx = \dfrac{P\uparrow \max + P\uparrow \min}{2}$

Under the bottom line of column 6 we sum up all marginal bucket-specific approximate values of P'↑ and obtain the overall value of P↑, approx.

The denominator of Score Potential in the case of this gains chart is calculated as 10+9.9+9.8+9.7=39,4 or according to

(9) Eq. 9: $\quad P\downarrow = BNo^{\max} * R(X) * X + \left(-\dfrac{\#B}{\#X}\right) * \dfrac{(R(X) * X - 1) + R(X) * X}{2}$

Further Columns: column 7 illustrates that the Score Potential is cut-off independent as it does not add up to 100% as is the case with the BenI in col. 9. Column 7 calculates the marginal value of the Score Potential per bucket by dividing the marginal value of P'↑ by P'↓ per bucket. Col. 10 gives us the maximum of BenI according to equation 2 above and col. 11 the ratio of BenI / BenI max per bucket.

The final approximation of Score Potential is now calculated as

(10) Eq. 10: $PoP_{approx} = \dfrac{P\uparrow approx}{P\downarrow} = 74\%$

At first glance, this Score Potential value reveals that 74 % of the best possible model has been achieved. Having that information it is easy to set a stretched target of e.g 80%. An additional advantage delivered by the Score Potential becomes clear in the following exercise: Let us assume the model's forecast characteristic changes within consistent cut-off and shifts all 4 responders into the best bucket no. 10. This alteration changes the Score Potential from 74% to 97% and even to 100 % if all responders are at the top of bucket no. 10 (consequently the P↑ max value in col. 3 is used as numerator in equation 10).

The Score Potential-range between 75% - 100% besides a constant BenI is shown in the diagram below. The hatched area between the changing Score Potential curve and stable BenI line illustrates the room for improvement indicated by the Score Potential but not by the traditional benefit index BenI.

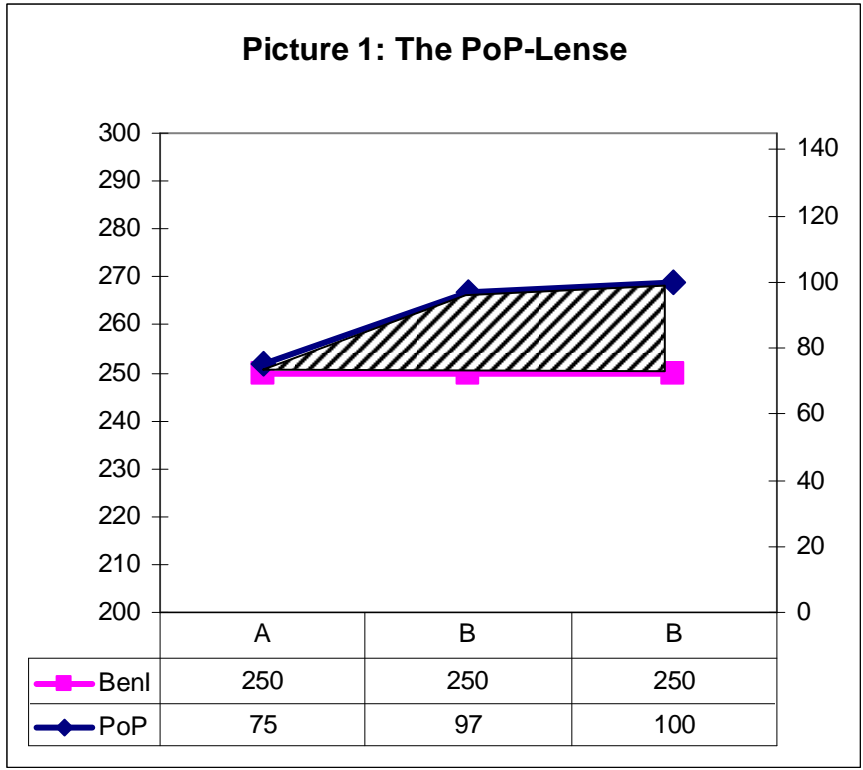

Picture 1: The PoP-Lense

|  | A | B | B |
|---|---|---|---|
| BenI | 250 | 250 | 250 |
| PoP | 75 | 97 | 100 |

Performing an effort-intensive manual model review one could observe the three different values of Score Potential (74%, 97%, 100%) by instantly looking at 3 gains-charts, however it is not always recommended to do that for the following reason. The more score models a direct marketing business churns out, the more costly the manual review of different gains charts. Concurrently the more benefit is spent by a computerized review. This leads to a further advantage of the Score Potential: to serve as an "all-purpose" measure for a programmed model evaluation. Especially high-end direct marketing businesses which produce lots of score models will profit from a Score Potential-based computerized evaluation. On the one hand this saves costs of a labor-intensive manual review. Moreover, it will re-focus human brainpower to all other creative areas than thinking about the quality of models: nobody needs to think about alternative values of a traditional benefit index, or what they mean for different cut-offs. Likewise nobody will need to think about different method-driven test statistics, how to compare them in case of different statistical approaches and what they mean for the resulting benefit under concrete economical circumstances. The more a high-end direct marketing business is characterized by high promotion frequencies, a broad variety of sales promotion, big customer potentials and large databases with lots of powerful business intelligence data, the more weight the advantage of a Score Potential based computerized evaluation will carry.

# 4    Summary: Benefit of the potential-oriented performance measure Score Potential

Based on the above considerations we formulate the following thesis. The potential-oriented performance measure Score Potential basically presents three advantages versus traditional benefit indices:

the first is of a psychological nature. Good data analysis is not just driven by a sound analytical competence and craftsmanship but also by an inquisitive and ambitious mindset. Benefit indices do not foster that attitude. They lull analysts by indicating an apparently satisfying gain of score models. This applies in particular if cut-offs are traditionally set and score gains conform to ditto traditional expected gains. The Score Potential does not have that handicap of a benefit index. By measuring the distance to an optimum, it raises the question: is there an opportunity to get better? Therefore it implies a stretch goal, which tends to lead to better effort and performance of analysts.

the second advantage is that Score Potential highlights rooms for improvements, which have to stay in the dark for usual benefit indices. It has been exemplified (picture 1) that Score Potential detects different model qualities where traditional benefit indices do not show any change.

the third advantage of the Score Potential is the ability to serve as a measure for a computerized model evaluation. Businesses churning out score models galore may profit from the advantage of Score Potential-based programmed model evaluation: cost savings & re-focus on human creativity.